\documentclass[12pt]{iopart}
\usepackage{graphicx}
\usepackage{hyperref}
\hypersetup{
    unicode=false,          
    pdftoolbar=true,        
    pdfmenubar=true,        
    pdffitwindow=false,     
    pdfstartview={FitH},    
    pdftitle={Recent Advances in the Numerical Simulations of Binary Black Holes},    
    pdfauthor={Pedro Marronetti},     
    pdfsubject={Subject},   
    pdfcreator={Creator},   
    pdfproducer={Producer}, 
    pdfkeywords={numerical relativity: general relativity- numerical simulations - astrophysics - binary black holes - gravitational waves}, 
    pdfnewwindow=true,      
    colorlinks=true,       
    linkcolor=red,          
    citecolor=blue,        
    filecolor=magenta,      
    urlcolor=cyan           
}

\newcommand\ba{\begin{eqnarray}}
\newcommand\ea{\end{eqnarray}}
\newcommand\et{{\it et al.~}}

\begin{document}

\title[Binary Black Holes: Recent Advances]
       {Recent Advances in the Numerical Simulations of Binary Black Holes}

\author{Pedro Marronetti$^1$, 
Wolfgang Tichy$^1$}

\address{$^1$ Physics Department, Florida Atlantic University, Boca Raton, FL 33431-0991}

\ead{pmarrone@fau.edu} 

\begin{abstract}
Since the breakthrough papers from 2005/2006, the field of numerical relativity has experienced a growth spurt that took the two-body problem in general relativity from the category of ``really-hard-problems" to the realm of ``things-we-know-how-to-do". Simulations of binary black holes in
circular orbits, the holy grail of numerical relativity, are now tractable problems that lead to some of the most spectacular results
in general relativity in recent years. We cover here some of the latest
achievements and highlight the field's next challenges.
\end{abstract}



\section{Introduction}
\label{introduction}

Whether black holes ({\bf BH}), defined as a family of solutions to the vacuum Einstein field equations that present a one-way membrane (i.e., the event horizon) that causally separates two regions of spacetime, exist or not was a matter of speculation for the best part of the 20th century. In the past decade, however, astronomical observations placed them as the most promising models for objects detected in X-ray binaries (with sizes of a few to tens of solar masses) and for the supermassive entities at the center of most galaxies (with up to billions of solar masses) \cite{Narayan:2005ie}. While they can be fully specified by their mass, angular momentum and charge, only the former two are of astrophysical relevance since charged BHs are quickly neutralized by free charges found in their vicinity (i.e, from accretion disks, interstellar plasma, etc.).

While individual BHs are extremely interesting objects on their own, when they couple in binary black holes ({\bf BBH}) they become more so by forming one of the most promising sources for gravitational wave detectors \cite{Cutler92}. The two-body problem in general relativity is unsolved in the sense that in this framework we do not have solutions analogous to the Keplerian curves of Newtonian gravity. Due to the emission of gravitational waves, the binary loses energy and angular momentum tightening the orbit until finally the two holes merge to form a single larger distorted BH. This still emits gravitational waves until it reaches the state of a stationary rotating BH. Until recently, the most detailed predictions existed only in two disjoint regimes where approximation schemes could be used. As long as the two objects are far apart, post-Newtonian calculations \cite{Futamase:2007zz} can give highly accurate approximations for the orbital motion and the gravitational waves emitted by the binary. When the two BHs get closer, the post-Newtonian expansion becomes more and more inaccurate and eventually breaks down. Later in the evolution of the binary when the two BHs have already merged, the close limit approximation  \cite{Price94a, Pullin:1999rg, Khanna99a} can be used to model the ring-down of the remaining single BH. However, in the intermediate region that covers the last couple of orbits and up until the point when a single BH forms, the full non-linear Einstein equations have to be solved and computer simulations are used to obtain numerical answers \cite{Pretorius:2007nq}. Such simulations are a hard and challenging problem.
A problem that with the first generation of interferometric gravitational wave detectors fully operational and on their way to their second generation, has become increasingly important. As sensitive as these instruments are, it seems likely that experimental detection will hinge on detailed and accurate theoretical predictions. Such predictions would help distinguish the weak signature of real astrophysics among the various types of noise known to be present in the apparatus. 

Enter numerical relativity. The field of numerical relativity is relatively young since its development has been closely paired to the growth of the power and availability of computers. While some pioneer work was done as early as the mid 1960s, it wasn't until the 1970s that the first results of BBH spacetime evolution were obtained \cite{Smarr76, Eppley75}. These corresponded to axi-symmetric BH collisions. Researchers soon found that the situation gets much more difficult when dealing with problems without a high degree of symmetry as in the case of BBH in semi-circular orbits. To say that these simulations are a hard and challenging problem is a gross understatement. Until recently they tended to fail after a very short time due to instabilities which resulted in exponentially growing run-away solutions. Fortunately, 
since the breakthrough papers of Pretorius \cite{Pretorius:2004jg}, Campanelli \et \cite{Campanelli:2005dd} and Baker \et \cite{Baker:2005vv}
tremendous progress has been achieved and it is now possible to evolve BBHs through many orbits and the subsequent merger and ringdown phases. We will cover here some of the most spectacular discoveries regarding BBH made through the solution of the full set of Einstein field equations.  Many of the foundations of the subjects touched here are described in the excellent books by M. Alcubierre \cite{Alcubierre_book} and T. Baumgarte and S. Shapiro \cite{Baumgarte_Shapiro_book}. Quantities in this article are expressed Geometrized units, where $G=c=1$.

\section{Binary Black Holes: Recent Developments and Discoveries}

\subsection{Orbital Hang-up and Naked Singularities}

\begin{figure}
\begin{center}
\includegraphics[scale=0.3]{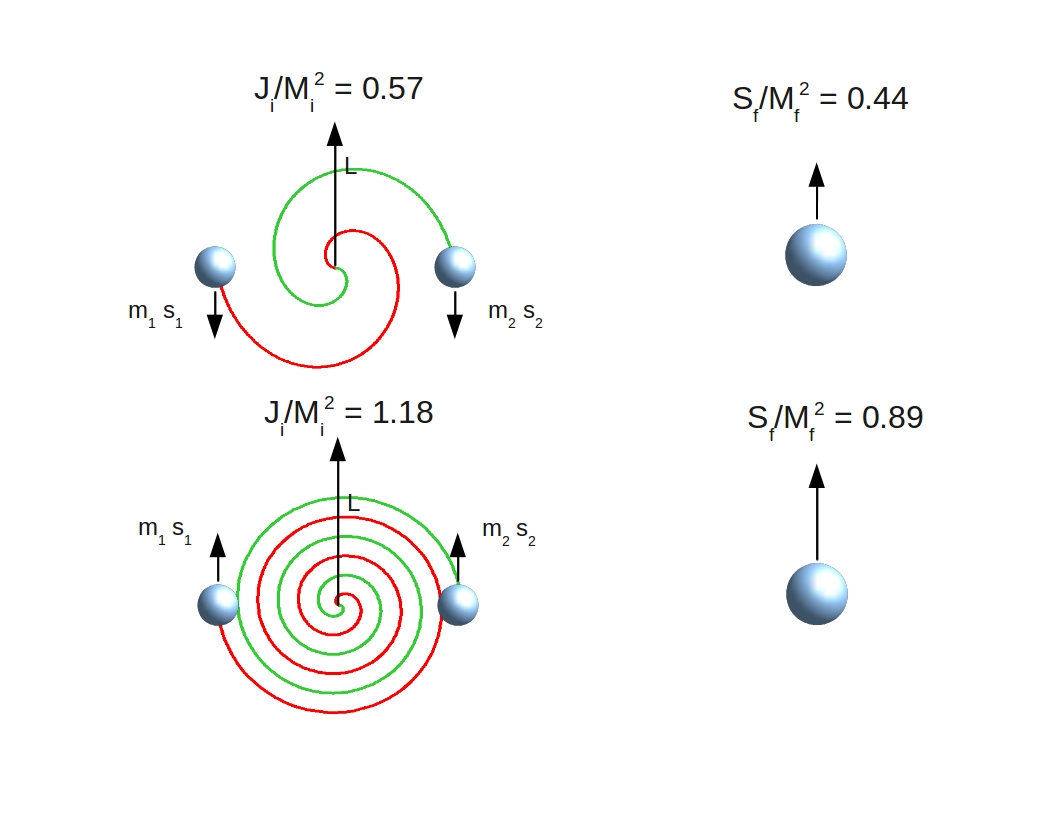}
\caption{At the top (right) we have a BBH system with components whose spins are anti-aligned with the orbital angular momentum, resulting in a total angular momentum less than critical. At the bottom (right) the same BHs are flipped to align the spins with the orbital angular momentum, resulting in a total angular momentum higher that critical. Numerical simulations show that the lower system will take a longer time to merge, emitting the excess of angular momentum that would have otherwise produce a naked singularity. The numerical values are those of Campanelli \et \cite{Campanelli:2006uy}.}
\label{spin_hangup}
\end{center}
\end{figure}

  Relativists have speculated for almost a century about the possibility of finding naked singularities (i.e., those that do not hide behind event horizons). Penrose in his famous {\it Cosmic Censorship} conjecture stated that no naked singularities exist apart from the one associated with the Big Bang. Choptuik showed in 1997 \cite{Choptuik:1996yg} that under certain very particular set of conditions ``non-generic" naked singularities could be created \footnote{Simply put, ``non-generic" singularities are such that arbitrarily small perturbations would send them back inside event horizons.}. Thus, the search for mechanisms (natural or artificial) that would lead to the formation of naked singularities continues. Different astrophysical scenarios such as rotating BHs accreting mass-energy, hypermassive neutron star collapse, hypernovae, high-speed BH collisions, etc., have been studied with the goal of forming stationary compact objects with a normalized angular momentum higher than critical ($S/M^2 > 1$) which would indicate the presence of a naked singularity. So far all these studies have lead to negative answers. One particular scenario is that of BBH mergers whose total normalized angular momentum at some point during the inspiral is larger than the critical value. Campanelli \et \cite{Campanelli:2006uy} studied first this situation by considering two BBH systems composed by two equal-mass BHs positioned at the same separation and with dimensionless spin of magnitude $s/m^2=0.757$. However, in one case the spins were aligned with the orbital angular momentum (leading to an initial angular momentum of $J_i/M_i^2=1.18$) and in the other they were anti-aligned (now with $J_i/M_i^2=0.57$). Figure \ref{spin_hangup} shows these configurations.  Their study showed that the dynamics of the final orbits are significantly affected by the orientation of the spins. The aligned case {\it required} three orbits before the merger, radiating $\sim 7\%$ 
of the total energy, while the anti-aligned case radiated only $\sim 2\%$, merging in less than one orbit. In both cases the spin of the final BH was less than critical, indicating that forming naked singularities by the way of merging BHs may not be possible. This conclusion has been confirmed so far by several groups that have studied mergers with different masses and spins of arbitrary magnitude and direction. In particular, the formulas that predict the final mass and spin of a BBH merger described in Section \ref{final_mass_spin}, albeit approximate and only accurate for BHs with comparable mass, fail to predict a region of parameter space for which BHs with higher-than-critical spin could be formed.

\subsection{Black Hole Kicks and Anti-kicks}

\begin{figure}
\begin{center}
\includegraphics[scale=1.0]{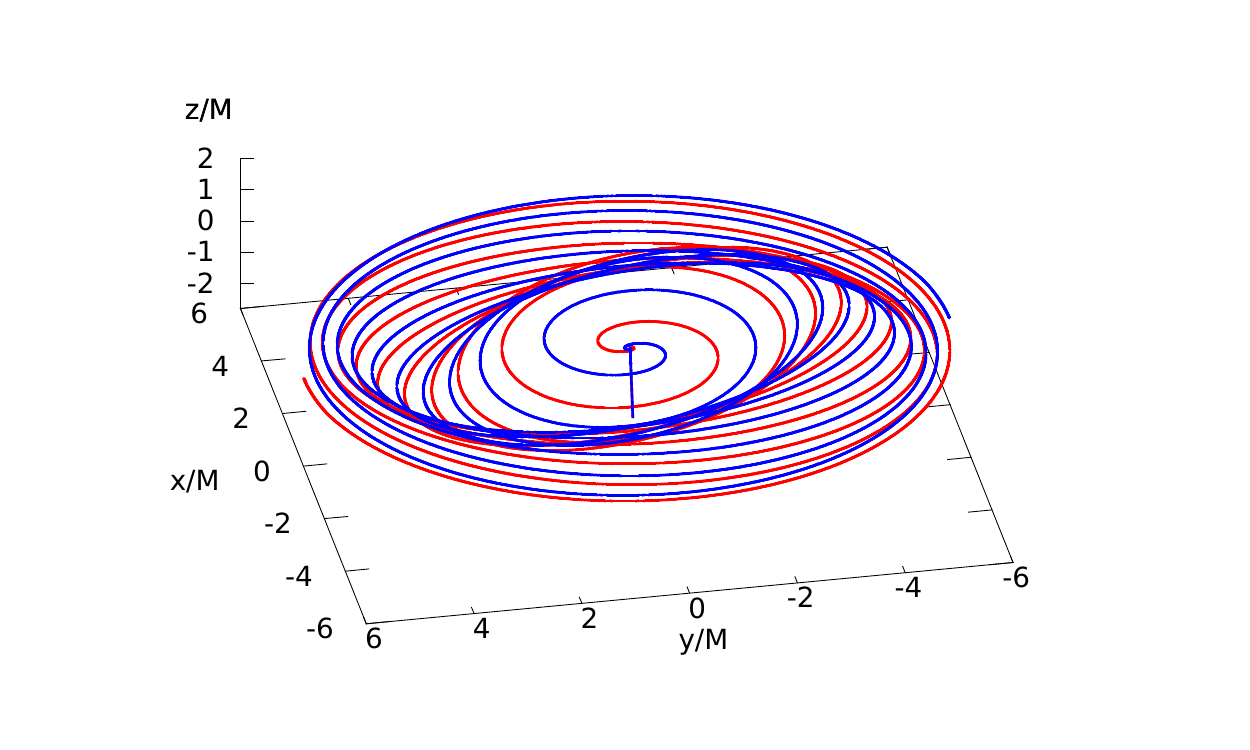}
\caption{Trajectories of the black holes in a BBH simulation. In this case, the individual BHs have arbitrary spins which force the orbital plane to precess and the final BH to acquire a recoil velocity (``kick") represented by the straight line anti-parallel to the z-axis.}
\label{bbh_kick}
\end{center}
\end{figure}

With the exception of highly symmetric
situations such as head-on collisions of equal mass non-spinning BHs,
generic BBH will radiate linear and angular momentum.  This effect has
important consequences for astrophysics.  It means that any BBH with unequal
masses will acquire a kick due to gravitational radiation during the
inspiral and merger of the system.  The magnitude of the resulting recoil is
important in a variety of astrophysical scenarios, such as the cosmological
evolution of supermassive BHs or the growth and retention of 
intermediate-mass BHs in dense stellar clusters.  For a binary in almost
circular orbit, the direction of the instantaneous linear momentum flux
rotates in the orbital plane with the angular velocity of the system.  Thus,
when the binary goes through one orbital period, the average linear momentum
flux will be close to zero.  The only net effect comes from the fact that
the inspiral orbits are not perfect circles.  Most of the kick is
accumulated during the last orbit and subsequent plunge of the two holes,
when the motion is no longer quasi-circular and the averaged linear momentum
flux is much larger.  Several analytical estimates of the kick velocity have
been published in recent years ~\cite{Wiseman:1992dv, Favata:2004wz,
Blanchet:2005rj, Damour:2006tr}.  All these estimates have been derived
using approximations (such as post-Newtonian theory) which break down during
the last orbit missing the strongest contribution of the gravitational radiation to the kick velocity. On the other hand great progress has been made in the past couple of years with full GR calculations (see \cite{Baumgarte_Shapiro_book} and references therein). It is now possible to calculate the kick velocity for any initial configuration using numerical simulations.

For non-spinning BHs the maximum kick velocity occurs for a mass
ratio of $q \equiv m_1/m_2=2.77$ and has a magnitude of 175~km/s~\cite{Gonzalez:2006md}. 
For equal mass BHs a kick occurs only if the initial BHs are
spinning.  If the spins of the initial BHs are aligned or
anti-aligned with the orbital angular momentum the maximum kick velocity has
a magnitude of 448~km/s~\cite{Pollney:2007ss} directed in the orbital
plane.  This maximum occurs for maximally spinning BHs with
spins that are anti-aligned with each other.
However, much larger kicks are possible for equal-mass
binaries with anti-aligned initial spins in the orbital plane.  Simulations
have yielded kicks as high as 2500~km/s~\cite{Gonzalez:2007hi}. 
Extrapolations from these simulations by Campanelli et
al.~\cite{Campanelli:2007cg} predict kick velocities as high as 
4000~km/s. 
Kicks on the order of $1000$~km/s seem to be a
generic feature even if the initial spins are not exactly in the orbital
plane and not exactly anti-aligned~\cite{Tichy:2007hk}. 
Some of the latter kicks are so high that the
final BH would escape both from dwarf elliptical and
spheroidal galaxies (with typical escapes velocities of below 300~km/s) and
from giant elliptical galaxies (2000~km/s)~\cite{Merritt:2004xa}.  The
question thus arises how often such high kick velocities occur.
After all, most galaxies that have undergone mergers seem to retain
supermassive BHs at their centers. 
The most likely answer~\cite{Bogdanovic:2007hp} is
that in real galaxy mergers torques from accreting gas align 
the spins with the orbital angular momentum, 
reducing the maximum velocity to a few hundreds km/s.

If one monitors the velocity of the center of mass during the inspiral and
merger, one finds that the velocity continuously changes direction while
slowly growing in magnitude during the inspiral. The final black
hole directly after merger is at first distorted and still radiates while
settling down to a more symmetrical stationary state. In many cases this 
radiation is directed such that center of mass speed decreases again.
This phenomenon is known as the
anti-kick~\cite{Schnittman:2007ij,LeTiec:2009yg,Rezzolla:2010df}.

\subsection{Spin Flips}
\label{spin_flips}

\begin{figure}
\begin{center}
\includegraphics[scale=0.3]{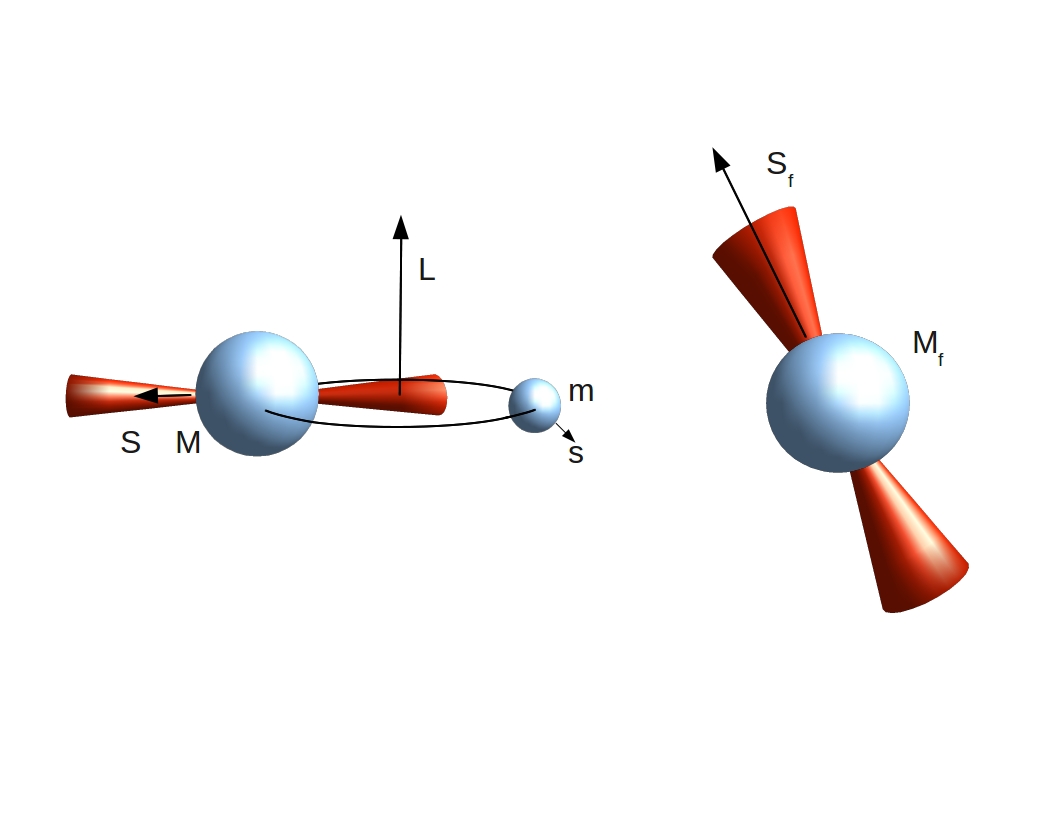}
\caption{Binary black hole merger model of the change in the direction of the spin observed in some supermassive BHs. Two galaxies merge forming a BBH (left) that in general will have BH spins and orbital angular momentum in arbitrary directions. The black hole of mass $M$ is surrounded by an accretion disk that powers the jets that are originally aligned with the spin $\vec{S}$. After the merger (right) the newly formed BH will have a new spin direction $\vec{S}_f$ that, if the original BHs are of comparable mass, will be mostly dictated by the pre-merger orbital angular momentum $\vec{L}$.}
\label{spin_flip}
\end{center}
\end{figure}

Black holes with masses ranging from millions to several billions of solar masses are colloquially known as {\it supermassive} and are found at the center of most galaxies. Of particular interest are those found in quasars and in galaxies with Active Galactic Nuclei (AGN) since they are believed to be the engine behind powerful jets that could reach kiloparsec scales. These jets are launched perpendicularly to the inner part of an accretion disk that surrounds the central BH and, barring a highly dynamical situation, they should be aligned with the BH axis of rotation (Bardeen-Petterson effect). Radio observations have shown that some galaxies have jets that at some point in their past changed abruptly directions. A class known as X-shaped radio galaxies \cite{Parma:1985, Leahy:1992} show the most dramatic cases, where the radio lobes in the inner region (younger) are shifted by almost ninety degrees with respect to those in the outer zones (older). 
This sudden change of jet orientation is associated with a change in the direction of the spin of the BH. Several theoretical models have been proposed to explain these phenomena that involve backflow from the main lobes and instabilities of the accretion disk. However, the most widely accepted explanation is the model that invokes a BBH merger. The idea is that the original AGN galaxy merges with a neighboring galaxy with its own central BH and both holes become gravitationally bound forming a binary. If the BHs are of comparable mass most of the angular momentum of the binary will reside in the orbital component of the angular momentum and, in general, this component will not be aligned with the spin of the AGN black hole. After the merger, the remnant resulting from the binary merger will have a final spin that will depend of the contribution of the binary's orbital angular momentum, the two pre-merger BH spins and the amount of angular momentum lost to gravitational waves. Figure \ref{spin_flip} shows a diagram that reflects this 
interchange. This model was originally posed by Merritt and Ekers \cite{Merritt:2002hc} and Zier and Biermann \cite{Zier:2002gr} and later confirmed by the full relativistic simulations of Campanelli \et \cite{Campanelli:2006fy}. There it was shown that the change in direction between the spin of the larger hole and the spin of the final BH can exceed $90$ degrees, consistent with observations of radio X-shaped patterns. They also exemplify the astrophysically relevant case of gradual semi-periodical shifts in the jet direction. Changes of this class are related to the precession of the BH spin during the inspiral and have also been observed in radio \cite{Komossa:2006}.

\subsection{Mass and Spin of the Final Black Hole}
\label{final_mass_spin}

The merger of a BBH will form a final larger BH of mass $M_f$ and spin $S_f$.
Thus, the initial state is described by eight parameters:
the mass ratio $q$, the spin components of the two pre-merger BHs
and the binary's angular velocity $\omega$. 
Here $\omega$ specifies the starting point of a possibly very long 
inspiral trajectory. After the merger the final BH is characterized 
by seven parameters, the final mass $M_f$, the spin $\vec{S_f}$ and 
the kick velocity $\vec{k}$. Predicting the final mass and spin 
from the initial parameters is of great importance in many astrophysical
merger scenarios.

Several groups have developed formulas that attempt to predict the
final spin of the merger. The analytic estimate of Buonanno et al.~\cite{Buonanno:2007sv} 
can give the final spin magnitude to within few percent with   
larger deviations for spins close to anti-alignment.    
There are also formulas which have been derived by fitting to results
from full numerical
simulations ~\cite{Campanelli:2006uy,Tichy:2008du,Barausse:2009uz}.
The approach by Campanelli et al.~\cite{Campanelli:2006uy} gives only
the magnitude of the final spin. This approach was later improved
by Lousto et al.~\cite{Lousto:2009mf} who give equations for all final spin
components, the final mass and the kick velocity. However, the formulas
in~\cite{Lousto:2009mf} still depend on the infall direction at merger, i.e.
on a quantity that is not known without performing numerical
simulations. The approach by Barausse and Rezzolla~\cite{Barausse:2009uz}
gives only the final spin, but all fitting coefficients
are explicitly given.
The formulas by Tichy and Marronetti~\cite{Tichy:2008du} predict the
the final spin vector as well as the final mass, and also do not contain
any unknown coefficients. Comparisons of these 
formulas~\cite{Barausse:2009uz,Kesden:2010yp} seem to indicate
that both the approaches in~\cite{Tichy:2008du} and ~\cite{Barausse:2009uz}
give similar results and have comparable errors. However, these
comparisons were done for low or moderate initial spins.
Lovelace \et \cite{Lovelace:2010ne} 
have performed two numerical simulations that test these different formulas
for the case of two initial BHs with very high spin (dimensionless spin
magnitudes above $0.94$). In these two cases the formula of
Tichy and Marronetti~\cite{Tichy:2008du} predicts a final spin value that
is closest to the true numerical answer. However, more high spin
cases with mass ratios different from one need to be studied before
one can draw firm conclusions.

These formulas are useful in astrophysical models where BBH
mergers are important. E.g. one can study how successive mergers of 
 BBH \cite{Tichy:2008du} influence the final spin.
Using the formula in ~\cite{Tichy:2008du} for the final mass both
O'Neill \et ~\cite{O'Neill:2008dg} and Rossi \et ~\cite{Rossi:2009nk}
have considered how the mass loss due to merger affects a circumbinary disk.
O'Neill et al. argue that the reduction of luminosity caused by the retreat of the
inner edge of the disk following mass loss could be detectable.
The same scenario is investigated by
Rossi \et ~\cite{Rossi:2009nk} who cast doubt on its detectability, unless
the final BH receives a substantial kick directed close to the disk
plane.

\section{Gravitational Wave Observatories and Numerical Relativity}

As the latest generation of gravitational wave detectors comes on-line, the problem of faithfully simulating the evolution of binary systems of compact objects has become increasingly important.  The detectors (NSF's LIGO \cite{LIGO_web}, VIRGO \cite{VIRGO_web}, TAMA \cite{TAMA_web}, GEO600 \cite{GEO_web}) use laser interferometry to measure the tiny strains associated with passing gravitational waves  \cite{Schutz99}, offering much higher sensitivity than previous experiments. In addition, new detectors are under planning and construction stages such as the Einstein Telescope \cite{ET_web}, Indigo \cite{Indigo_web} and DECIGO \cite{Kawamura:2011}. As we mentioned in Section  \ref{introduction}, due to the very low signal-to-noise ratio LIGO/VIRGO ({\bf L/V}) data analysts employ a method known as {\it matched filtering} in order to boost the chances of detectability. This technique requires a priori knowledge of the shape of a given signal which is compared with observational data to assess the probability of its presence amidst the noise. LIGO/VIRGO scientists and collaborators have created a bank of templates based on post-Newtonian methods that cover the early stages of the binary life (inspiral) which have been successfully incorporated to the data analysis pipelines used in all the scientific runs corresponding to the initial stage of the detectors \footnote{Since October 2010, LIGO has been taken off-line to undergo the upgrades leading to the next generation observatory Advanced LIGO.}. 

Now that the numerically relativity community has achieved the maturity level needed for BBH simulations, many groups in the field (including ours) have clustered around several collaborative projects that directly involve L/V data analysts. Binary black holes in quasi-circular orbits can be described by eight parameters (see Section \ref{final_mass_spin}). Additionally, in order to remove any orbital eccentricity \cite{Pfeiffer:2007yz,Boyle:2007ft,Husa:2007rh,Walther:2009ng,Tichy:2010qa} that might be present in the initial data, a careful determination of the initial tangential momentum ensues adding an extra parameter \footnote{Gravitational radiation has a circularizing effect on the BBH orbits. Any eccentricity found in the BBH at the moment of formation is expected to have radiated away by the time of the merger.}. Covering a nine-dimensional parameter space with month-long simulations is such a daunting task that is clear that the L/V needs can only be met
  (at least in the short and intermediate term) by a close collaboration between numerical and analytical relativists. This common understanding in the relativity community has given birth to the NRAR (Numerical Relativity - Analytical Relativity) collaboration \cite{NRAR_web}. In addition to this effort, numerical groups around the world are also converging into a tightly knit collaboration with L/V data analysts that aims at incorporating numerically generated gravitational waveforms into the observatories data pipelines: the NINJA collaboration \cite{NINJA_web}.

\subsection{NINJA and NRAR Collaborations}

The Numerical Injection Analysis (NINJA) project started more than two years ago and consists of the largest scientific collaboration between numerical relativists and L/V data analysts to date. NINJA's goal is the smooth incorporation of BBH numerical templates into the observatories detection infrastructure. During the first stage of the collaboration the waveforms provided by the numerical groups were embedded in colored Gaussian noise and injected into the data analysis pipelines. After these results were published \cite{Aylott:2009ya_short, Aylott:2009tn_short}, plans for a follow-up project were drawn giving birth to NINJA-2. The goal of the second stage of the project is producing longer and more accurate gravitational waveforms by the numerical groups and executing more systematic tests, this time using real L/V data noise. Our team has been part of NINJA since its inception. NINJA-2 has set guidelines on the length and quality of the waveforms to be prepared by the numerical groups. One of them is that the simulations should cover at least ten waveform cycles (five BBH orbits) not including the short period at the beginning of the simulation when the signal is corrupted by the ``junk" radiation found in the initial data sets. Other restrictions are that the error accumulated in the amplitude and phase of the waveforms $l=2 ~m=2$ mode during the inspiral should not exceed $5\%$ and $0.5$ radians respectively.  

The Numerical Relativity - Analytical Relativity (NRAR) collaboration was kick-started at the end of 2009 and consists of a thrust by a large group (more than 50 members) of numerical and analytical relativists. The goal is to coordinate recent developments in the production of BBH waveforms in both numerical and analytical methods. One of the goals of the collaboration is the generation of a small survey of high-accuracy relatively long (20 cycles or more) numerical waveforms to be used in the calibration of analytical template banks. These can then be used to generate efficiently and faithfully tens of thousands of templates for L/V matched-filtering search algorithms. Like in the case of NINJA-2, the NRAR collaboration has also set guidelines on the length and quality of the waveforms. NRAR requires waveforms with at least twenty waveform cycles and an accumulated phase error not to exceed $0.05$ radians. In
  addition to these criteria, NRAR opted for minimizing the initial orbital eccentricity to make the signals more ``astrophysically realistic" (see Section \ref{final_mass_spin}). 

\begin{figure}
\begin{center}
\includegraphics[scale=0.4]{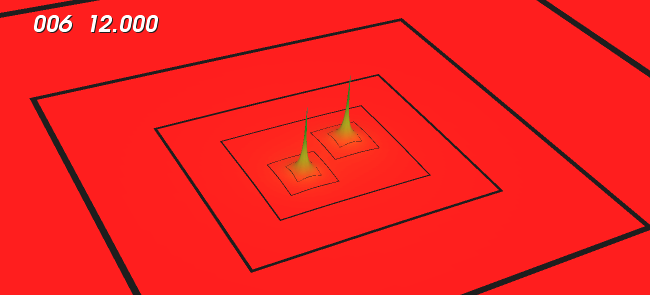}
\caption{ The solid black lines mark the boundary of each refinement level. Shown are four concentric {\it fixed} and two {\it moving} levels around each BH. Each level doubles the spatial resolution.}
\label{moving_boxes}
\end{center}
\end{figure}

\subsection{Characteristics of a Typical Binary Black Hole Simulation}
\label{sims}

\begin{figure}
\begin{center}
\includegraphics[scale=0.3]{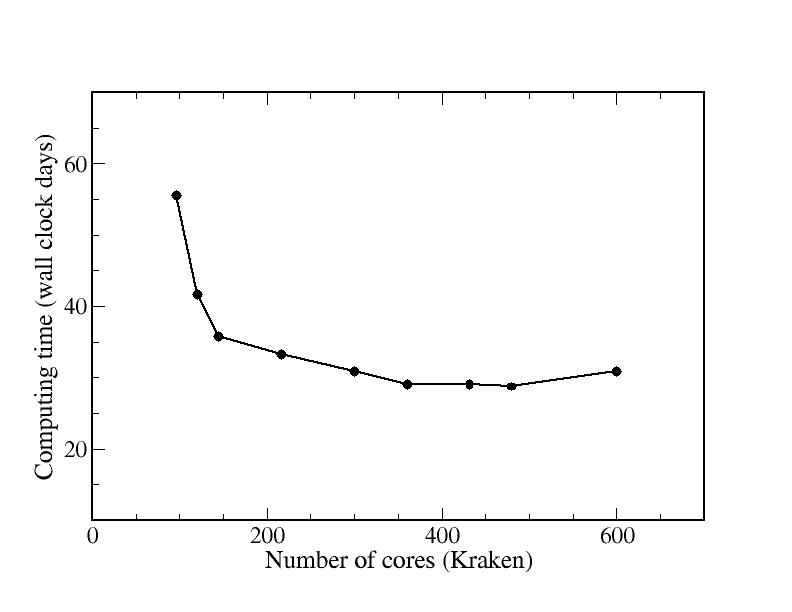}
\caption{ Computing time (measured in wall clock days) vs. number of cores for a full-grid BBH simulation such as the one described in Section \ref{sims}. Scaling was performed on NICS Cray XT-5 {\it Kraken}.}
\label{bam_scaling}
\end{center}
\end{figure}

In this section we describe some of the characteristics common to the most advanced simulations that are being produced by our group for the NINJA-2 and NRAR projects. The simulations are performed with the code BAM \cite{Bruegmann:2003aw,Brugmann:2008zz, Marronetti:2007ya, Marronetti:2007wz, Tichy:2007hk, Tichy:2008du} which evolves the gravitational fields using the BSSNOK formalism \cite{Nakamura87, Shibata95, Baumgarte:1998te} in the variation known as the ``moving punctures" method \cite{Campanelli:2005dd, Baker:2005vv}. BAM is based on a method of lines approach using sixth order finite differencing in space and explicit fourth order Runge-Kutta (RK) time stepping. The particulars of our numerical implementation can be found in \cite{Bruegmann:2006at, Marronetti:2007wz}.
The numerical domain is represented by a hierarchy of nested Cartesian boxes as shown in a simplified manner in Fig. \ref{moving_boxes}. It consists of $L+1$ levels of refinement, indexed by $l = 0, \ldots, L$. A refinement level consists of one or two Cartesian boxes with a constant grid-spacing $h_l = h_0/2^l$ on level $l$. Typical values are $L+1=12$ for the number of refinement levels, with the levels 0 through 5 each consisting of a single fixed box centered on the origin (the center of mass). On each of the finer levels (6 through $L$), we initially use two sets of moving boxes centered on each BH. When the BHs get close enough that two of these boxes start touching, they are replaced by a single box. The scaling of wall-clock time with number of processors is given in Fig. \ref{bam_scaling}.

\begin{figure}
\begin{center}
\includegraphics[scale=0.3]{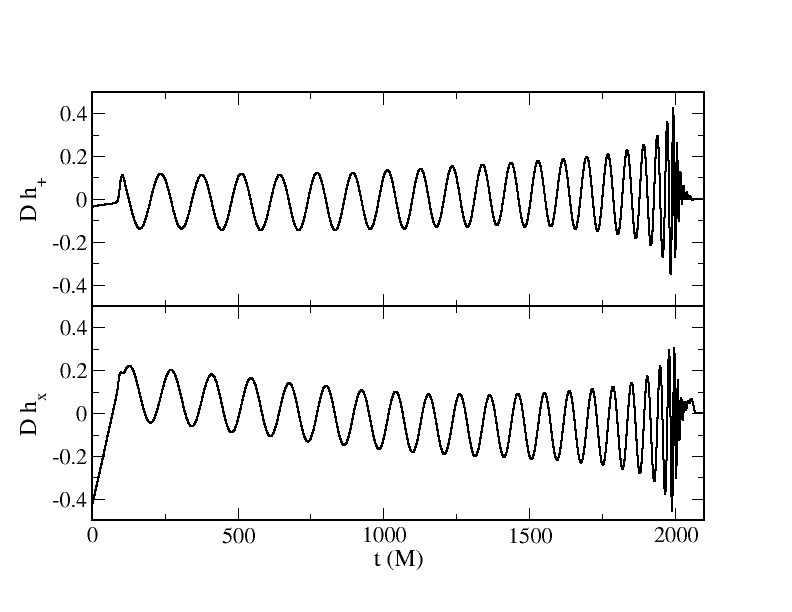}
\caption{{\it Left:} Gravitational wave strain with polarizations $h_+$ (top) and $h_\times$ (bottom) times the distance to the observer $D$ vs. time measured in units of total mass. The signal corresponds to an equal mass binary with spins of equal magnitude $s_1/m_1^2=s_2/m_2^2=0.6$ but arbitrary orientations.}
\label{hl2m2}
\end{center}
\end{figure}

\begin{figure}
\begin{center}
\includegraphics[scale=0.3]{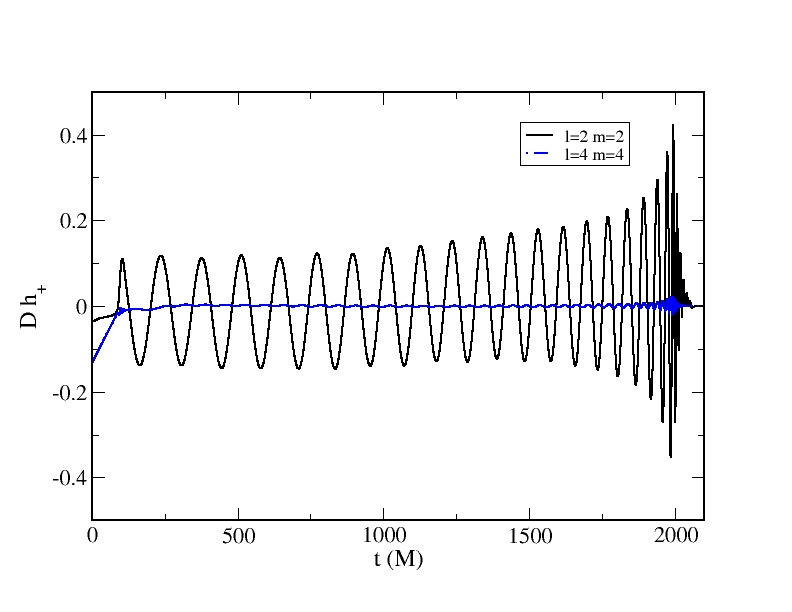}
\caption{{\it Left:} Gravitational wave strain with polarization $h_+$ times the distance to the observer $D$ vs. time measured in units of total mass. The curves shows the largest $m$ modes for the moments $l=2,~4$ (cases where $l=m$) of the simulations from Fig. \ref{hl2m2}. It can be seen that most of the power is emitted in the quadrupole ($l=2$) mode.}
\label{hl234}
\end{center}
\end{figure}

Figure \ref{bbh_kick} shows the trajectory of the individual BHs in an equal
mass binary.  The initial BH spins are 
$\vec{s_1}/m_1^2=(0.5196,0,0.3000)$ and
$\vec{s_2}/m_2^2=(0,0.5196,-0.3000)$ which have the magnitude
$s_1/m_1^2=s_2/m_2^2=0.6$.  These
parameters make the simulation fully three-dimensional: there are no
underlying symmetries that can be exploited for the sake of computational
efficiency.  The system is evolved for about $10$ orbits and the
gravitational waveforms corresponding to the dominant quadrupole mode
($l=m=2$) are shown in Fig.  \ref{hl2m2}.  Figure \ref{hl234} compares the
largest contribution for the multipoles with ($l=2$ and $4$),
highlighting the dominance of the quadrupole component.  In both graphs the time is expressed in units of $M$, the total mass of the system.  The merger produces a final BH with a total mass $M_f=0.952$ (from an initial fiducial mass $M_i=1.0$), with dimensionless spin of $S_f/M_f^2=0.704$ and a recoil velocity of $398~km/s$.

\section{Summary}

We have reviewed some of the most impressive achievements produced by the numerical solution of the Einstein field equations corresponding to BBH. A long path has been traveled in the few years since the community finally came to master the intricacies of these simulations. A much longer road awaits ahead. Currently, most groups that used to work on vacuum problems are making inroads into simulations with matter such as binaries with neutrons stars and white dwarfs, core-collapse supernova and BBH embedded in circumbinary disks. To the challenges inherent to dealing with the rich microphysics required by the latter simulations, we should add numerical challenges such as preparing the code for the next generation of peta/exa-scale platforms. While these problems are at least as challenging as the ones from the past, the results from these short years indicate that numerical relativity has a finally entered its golden age and, anticipating the detection of gravitational waves by the new generation of observatories, it couldn't happen at a better time.

\section{Acknowledgements}

This work was supported by NSF grant PHY-0855315. Computational
resources were provided by the Ranger cluster
at the Texas Advanced Computing Center (allocation
TG-PHY090095) and the Kraken cluster (allocation TG-PHY100051)
at the National Institute for Computational
Sciences.

\section*{References}

\bibliography{references}

\end{document}